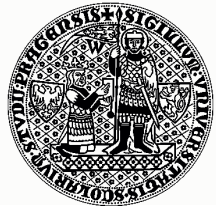

CHARLES UNIVERSITY PRAGUE

faculty of mathematics and physics

# Small Program For Fast Data Acquisition Of CAMAC And Other Devices On PC ISA bus.

P. Kodyš[1], J. Brož, Z. Doležal

Institute of Particle and Nuclear Physics
Charles University, Faculty of Mathematics and Physics
V Holešovičkách 2, CZ 180 00, Prague 8 – Troja, Czech Republic

## *Abstract*

A new data acquisition program for personal computers with Windows operating system has been built at Charles University van de Graaff laboratory. It allows fast data acquisition from CAMAC and other PC interface cards using interrupt or polling mode via several types of crate controllers, basic on-line analysis and visualization of acquired data in graphs and histograms. The program is written in combination of MSVC++ and ROOT macros and offers easy customization to individual user requirements via external configuration files, without need of recompilation.
In the presented paper a detailed description of the program concepts and features is given. Program was extensively tested with several combinations of hardware and software, and results of the tests are shown at the end.

---

[1] peter.kodys@mff.cuni.cz

## *1. Introduction*

The data acquisition (DAQ) system was built in the Van de Graaff accelerator laboratory of Institute of Particle and Nuclear Physics, Charles University, Prague for the purpose of the **N-D Polarization Experiment** to measure coincidence of four neutron detectors with one alpha detector. This means readout of approximately 20 CAMAC channels. An extension of DAQ to other PC devices is possible and easy, and it was tested by digital I/O interface PSD4848A. The main function of DAQ software is to move data from detectors and analog electronics via digitization part to the computer, save data to disk and display most important on-line results. Secondary requirements are to provide a user-friendly navigation that is understandable to the unqualified ground crew working in shifts.

Previous versions of our DAQ system used following platforms:

1. **Mainframe computers (SMEP)** with the programming language **Fortran**
2. PC's with **MS DOS** system and **Borland C++**
3. PC's with **MS Windows NT/2000** system and **MS Visual C++ 6.0** (without on-line monitoring plots)

All versions of DAQ software enabled secure saving of data in older systems (DOS) and old PC architecture (up to Pentium2) without any feedback about status and quality of acquired data. The version 3 of DAQ software has a non-trivial modification for CAMAC communication card and multi-parametrical spectra DAQ. Special conditions are available for on-line monitoring of values, their calculations and statistics. However, this version does not enable on-line histogram monitoring.

The objective is to guarantee data saving to more than one data storage for long-time experiments. Data will be saved with respect of free space on the disks. Users should be allowed to change many of the DAQ parameters and DAQ sequences in the configuration text files and run it without program compilation and rebuilds. Initialization sequences and self-monitoring utilities are included in main functions of the program. The multitasking and multithreading features of system are very useful for supported tasks, such as additional parallel off-line monitoring of DAQ status and data, backup of data files and connection to network.

These requirements served as background for development of the current DAQ software, which was built with respect to the state-of-the-art possibilities of PC hardware, operating systems and modern analytical programs.

## 2. Tasks and Conditions

Figure 1 shows standard scheme used for data acquisition in particle physics.

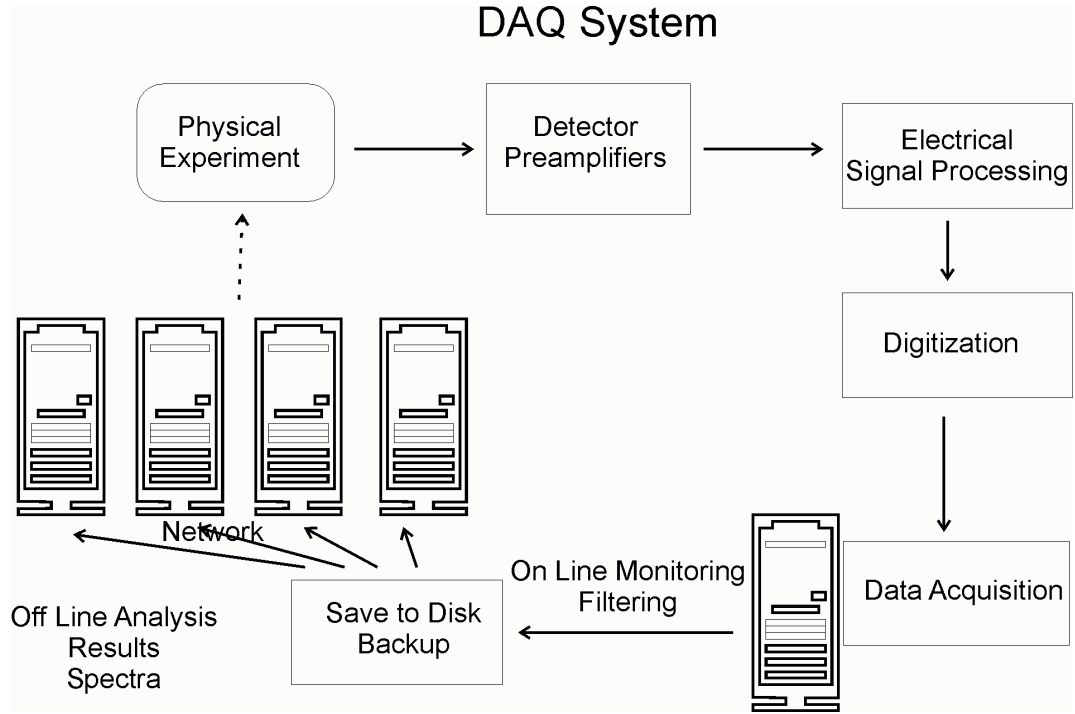

**Figure 1 Data acquisition scheme**

Physical experiments generate electrical signals, which are amplified and processed in analog electronics modules. Digitization units such as Analog-to-Digital and Time-to-Digital converters prepare signals for export to PC via e.g. CAMAC CC16 (Wiener [1]) or KK009 (Dubna [2]) interface or PC digital I/O interface cards such as PCD4848A. PC interface generates usually interrupt signal and interrupt in computer.

In the following the requirements and tasks which individual program parts should fulfill are listed.

### a. General

**Operating system** must ensure long time stability. Multitasking and multithreading system will be useful for support of parallel on-line monitoring of DAQ status and data, backup of data files and connection to network.
**Software** used must support quick DAQ processing, complex data processing – statistics, calculations. Very useful is graphical support of histogram and graph generation, programmable interface and possibility to customize to individual user requirements.

Chosen **program concepts** are summarized in the following points:

1. Use of trusted standards (used in physics community)
2. Program controlled by external configuration modules to enable non-programmers to change DAQ mode and some conditions
3. Data output format compatible with further off-line analysis inside MS Windows or LINUX (ROOT data format)
4. All important configuration values and actions must be saved automatically to a logbook file
5. Structuralism and comments in SW for clarity and continuity of SW operations and applications
6. User-friendly environment
7. Quality documentation and help support

### b. Data Acquisition

Data acquisition part should be based on several important aspects, such as work with PC interrupts generated by ISA bus, high communication speed with PC interface card (CAMAC PC16, PCD4848A), secure and efficient data saving, the maximum possible independence from other system activities, modularity to amend specialized SW modules changes at DAQ HW and separation of functional levels in SW for independence from used communication HW. Program must enable backup of data to more than one place for long time experiments. High rate of DAQ should enable to run on-line monitoring and slow control activities without freezing.

### c. On-line Monitoring and Slow Control Activities

On-line monitoring and slow control activities should never affect the DAQ speed. Required tasks for this part are:

1. Setting up and testing the measurement and DAQ devices
2. Display of:
    a. History of DAQ speed, error rates, statistics, etc.
    b. Cross coincidence tables
    c. one-dimensional (1D) histograms and two-dimensional (2D) scatter plots of spectra with special conditions
3. Preliminary background subtraction, peak areas.

## 3. Selected Tools

For the compatibility with earlier experience, the Windows operating system was preferred. Win NT and Win 2000 were selected for their stability.
Selected software should consider present status of our workgroup experience (see Introduction part) and modern trends. Finally, we have chosen following software for programming our DAQ system:

- **ROOT** (ROOT Development Team [3] ) - standard and complex data processing routines, including standard CERN libraries, full compatibility with C/C++, support of C/C++ quick modules, multiplatform SW: LINUX, MS Windows, easy transfer of ROOT data files and processing macros between different platforms.
- **Microsoft Visual C++6.0** (Microsoft) – fast communication driver support, fast procedures, specialized quick SW modules, prospective, dependable, quality design development, feasibility of special requirements for DAQ, fixed precise timer, computer interrupts, multitasking, multithreading, hardware driver control, dynamic libraries building

## 4. Program Structure Description

### a. Blocks and Encapsulation

Block arrangement of DAQ is shown in Figure 2. Data acquisition core part includes DAQ cycles connected via communication driver block to ISA bus driver and interrupt driver of PC card interface. Another part of software is connected to slow cycles, which are working in "soft" regime i.e. with respect to main DAQ cycle and basic data processing operation (calculations, saving, data transfer). Slow cycles ensure visualizations, histogram filling, control evaluation etc.

The individual layers (ISA bus driver, PC card interface driver and DAQ "hard" cycle) are strictly isolated from other software functions such as on-line monitoring. The ISA bus driver determines read/write/interrupt speed. PC card interface (for CAMAC or other special cards) driver sets type of communication hardware and its communication protocol.

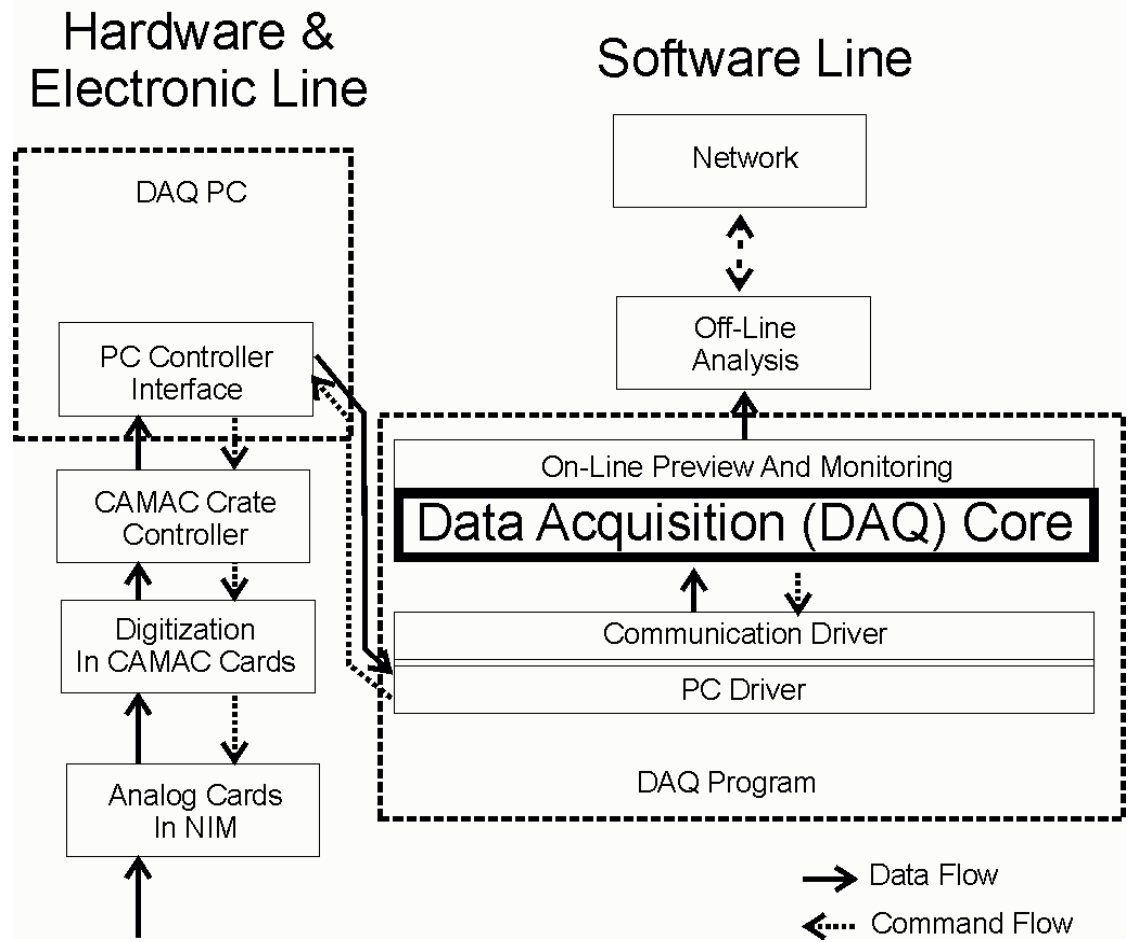

**Figure 2. DAQ software block architecture**

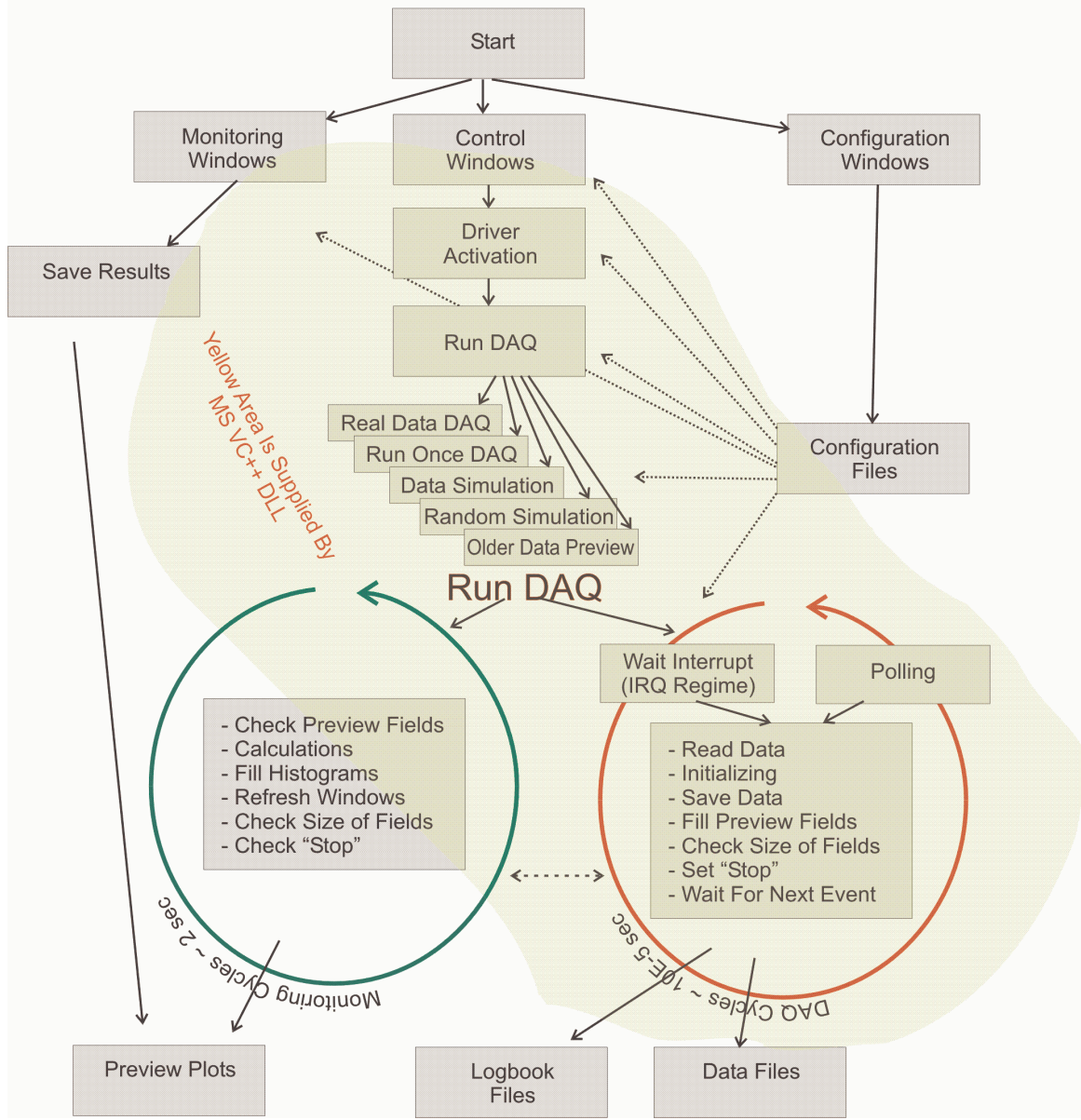

**Figure 3. Structure, data and command flowing in DAQ program**

### b. Data Acquisition Part

Data acquisition can be run in two modes: interrupt and polling mode. Interrupt mode uses standard PC interrupts generated by PC card. Polling uses quick cycle (Figure 3 red cycle) for looking to specific place in the memory and for checking if the system is ready for reading data event. CAMAC system generates automatically a signal that the system is ready for data reading (Look-At-Me – LAM), and this signal starts generation of interrupt in PC via PC card, or sets flag in PC card memory. Data acquisition was written in a C++ code strictly in C++ dynamic link library (dll). During DAQ, the program communicates with functions for data saving, control commands and statistics calculation in slow cycles. Special data files are shared with other parts of program, ROOT commands, and analytical part.

DAQ commands are divided to groups:
1. Device initialization (presets, configurations), which runs once in DAQ start and in slow periods.
2. Sequences for checking if device is ready for reading.
3. Sequences for data acquisition reading.
4. Sequences for reset of device for next event.
5. Sequences for self-calibration or device quality testing, if enabled. It is run in slow period.

Number of commands in all sequence parts is not limited.

### c. Statistics and Monitoring Part

Statistics calculations are part of quick DAQ cycle. Functions include cross comparisons, time dependent changes, efficiency, and area calculations. Results are filled to variables shared by visualization part. A high PC performance is very helpful at steering this part, especially if DAQ rate increases above 40 000 read/write per second.

The monitoring functions use "slow" (slow periodical) and "soft" cycles generated in ROOT environment (~seconds), which are run in a free DAQ time and have low priority. This means, the cycle is "frozen" in case of any problem or an overly high event rate.

### d. Visualization

During DAQ several on-line preview windows are displayed. ROOT tools were used for a full visualization, including a command line window for manual changes and instant interventions. On-line review windows show graphs of time-dependent changes, histograms and scatter plots. This part is built up on statistics and monitoring variables. The visualization was built in ROOT macro and some of the ROOT objects and variables are created and filled in a dynamic link library (dll). Not all library variables are shared with ROOT.

### e. Data Acquisition Control

The data acquisition control is organized in five levels:
1. Configuration external files used for control of main DAQ parameters and monitoring windows setting. Changes are made by editing of configuration files and reloading (restarting) program.
2. Top-level control using ROOT control tools, used for running of main actions, selection, preview and setting of DAQ parameters and different regimes, mirroring the control described in point 1. More details are in "Program Control Structure" section.
3. Special interactive procedures are given, which are set by navigation buttons within the window area, e.g. in spectra analysis: axis calibration, spectra detail editing and analysis. The navigation is user-interactive and partly saved.

4. The main DAQ cycle is user-controlled by a "STOP/PAUSE" window, which is an external object connected to a semaphore file for quick DAQ cycle.
5. Expert users may use the possibility to inspect and change any variable and object in ROOT environment and shared variables and objects in a mapped C++ dynamic link library. A command line window of ROOT is suitable for this control level.

### f. ROOT and MSVC++ Interconnection, Remarks on Compilation

Most important DAQ cycles and all quick processes were implemented in C++ dynamic link library (dll). The sharing, communication and control of this dll module is ensured by an interconnection with ROOT environment. The dynamic link library must also share ROOT objects by their mapping and compiling. The ROOT software structure fully supports this interconnection.

The quick DAQ cycle works with only C++ variables and functions without interconnection with ROOT shared variables and functions. Quick DAQ functions fill statistics buffers that are used for shared objects.

The monitoring part in slow DAQ cycles uses these shared objects for visualizations.

One remark on the process of program compilation may be of interest:
Since a quick DAQ cycle part does not share an environment with ROOT, the created list of shared variables and functions must exclude the group of quick DAQ variables and functions. For this a 3-step compilation process was used:
- Exclude quick DAQ variables and functions and save sources.
- Create shared interconnection file (automatically by ROOT program rootcint.exe).
- Include quick DAQ variables and functions and compilation and build dynamic link library.

### g. Program Control Structure

The program control structure depends on program application. Two basic applications are used: N-D experiment and spectra DAQ.
The program opens typically two basic windows and selected monitoring windows upon starting. Basic windows are the command line window and the main control menu.
Program control structure for N-D experiment is depicted in Figure 4.

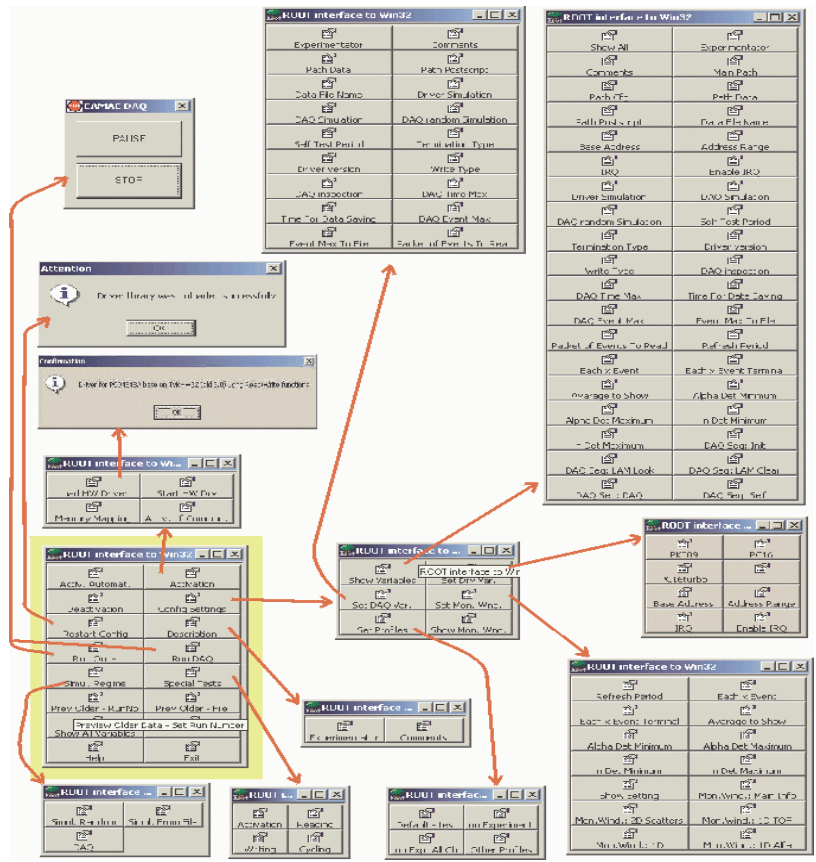

**Figure 4. Program control menu structure for N-D experiment. Starting menu is in yellow border.**

Main menu typically contains automatic or manual "step-by-step" activation, restart and deactivation of drivers and main functions of program, and configuration of variables. An important function, which is useful in the setting and calibration phase, is to run DAQ once. In a standard data acquisition process, control functions are transferred to the "STOP/PAUSE" window. Simulation and special functions (speed tests) are also in main menu. Help in html format is available.

### h. Program Configuration Structure

The basic configuration method of the program is via configuration files. This method guarantees a simple customization of the program to many different conditions and configurations. The program can be run *without compilation and rebuilds*.
Configuration files contain settings of paths, identifications, simulation flags, driver settings, DAQ setting, all DAQ commands (sequences), and monitoring and preview settings. Configuration of reading channels is also located in external file. Example of DAQ commands for setting NFA (standard form for CAMAC addressing) is in Example 1.

```
//NFA Settings: ***************************
// Sequences: Position N F A Value
// Position-> -1: write
// Position-> 0: read
// Position-> 0<: read & save
CM_DAQ_SEQ_DAQ0 1 19 2 0
CM_DAQ_SEQ_DAQ1 5 17 2 0
CM_DAQ_SEQ_DAQ2 6 17 2 1
```

**Example 1: Configuration file for format of NFA CAMAC addressing**

This format enables to write "Value" to address on an address range associated to a CAMAC card or to read BYTE or WORD from address to an ROOT ntuple[2]. Resulting values of a read event are put to specific position in a ntuple tree and saved to file in a ROOT standard format. The ntuple format is easily customizable to individual user requirements using external configuration file in form, as shown in the following Example 2.

```
ntuple|          |ntuple name |Wind |
Col   |Camac     |Channel     |Index|Comment
-----------------------------------------------
0      Event      EVT          0     Order in file
1      AD413A_0   AEN          1     AD413A card, channel 0
2      AD413A_1   AD_1         -1    AD413A card, channel 1
3      AD413A_2   AD_2         -1    AD413A card, channel 2
4      AD413A_3   AD_3         -1    AD413A card, channel 3
5      TDC414_0   TOF0         2     TDC414 card, channel 0
6      TDC414_1   PSD0         3     TDC414 card, channel 1
7      TDC414_2   TOF1         4     TDC414 card, channel 2
```

**Example 2: Configuration file for format of reading channels, where unsaved channels obtain "Wind Index" = "–1", saved channels have in this index the number of their position in ntuple, and ntuple column name is shown in "ntuple name channel".**

## *5. Program Applications*

### a. N-D Experiment

**Introduction**

Figure 5 shows experiment arrangement in our laboratory. The experiment's aim is to measure cross-sections of fast neutrons scattered off a deuteron target. Neutron beam is produced from the T(d,n)α reaction via an associated-particle method, i.e. α-particle ejected from the production target together with neutron is registered in coincidence with the neutron and serves as a tag of neutron. Anode and dynode signals from 4 neutron detectors (photomultipliers) are collected together with a signal from the α-particle detector. Individual signals are then used to define the coincidence, time delay between the signals and to discriminate background from γ's from the useful neutron signals. Both analog and time to digital converters are used, plus several scalers. In total about 20 signals are digitized in CAMAC modules. Further details about the system can be found at [4] and [5].

---

[2] ROOT ntuple is a matrix of columns of numbers in FLOAT format, each line of matrix is often called "event".

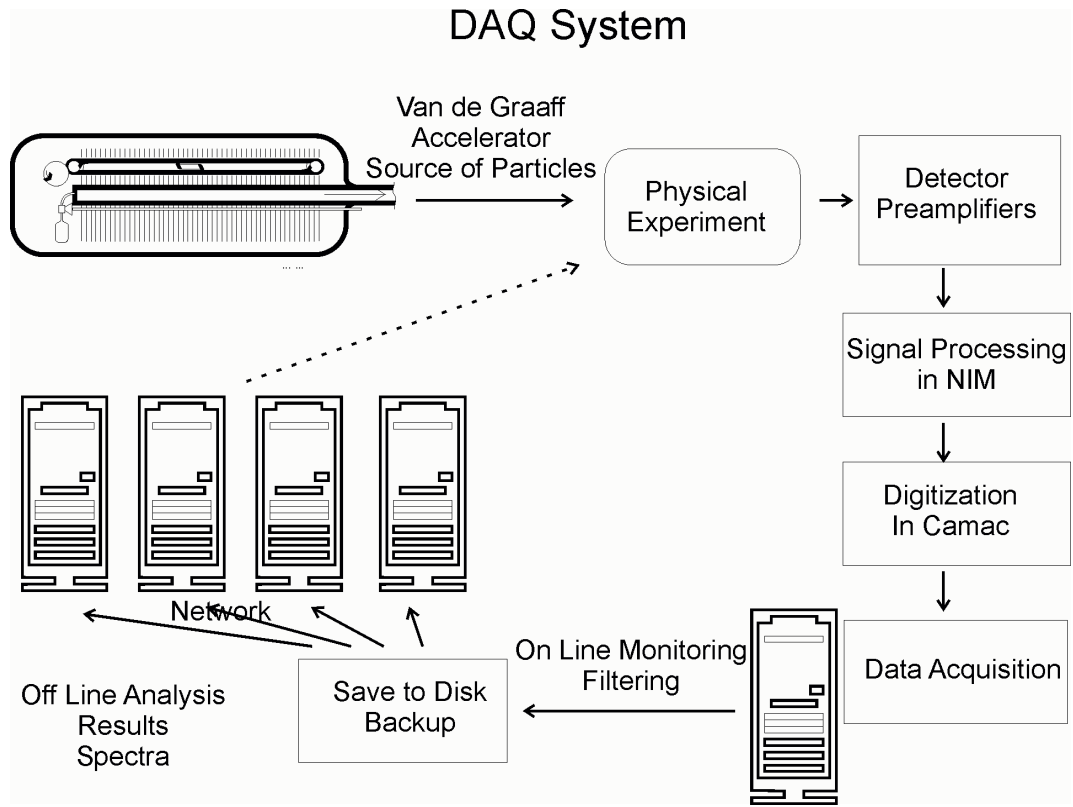

**Figure 5. N-D experiment scheme**

Electrical signal from detector preamplifiers is conveyed to NIM analog units for basic processing. CAMAC digitization units such as AD413A (EG&G Ortec), C414 (C.A.E.N.), AD2249A (LeCroy), NL2305 (Tesla) and NL2402 (Tesla) prepare signals for export to PC via CC16 (Wiener) or KK009 (Dubna) interface and generate CAMAC - LAM signal and interrupt in computer.

**Program Description**

The N-D experiment requires special monitoring windows with calculated values and 1D and 2D histograms. It is necessary for detecting possible DAQ problems – e.g. loosing of detector high voltage, burnout of target, loosing of primary particles or changes in magnetic fields. All windows contain buttons for clearing plots and for saving results to postscript file.

There are two windows showing calculation values: parameter tuning monitor (Figure 6, all windows are colored but in the paper the black-and-white screenshots are shown only) and standard on-line monitor (Figure 7). Parameter tuning monitor (Figure 6) shows frequencies of all coincidence combinations. Window shows tables of cross-references used in parameter tuning to optimal working point.

```
n-n Experiment - Tuning Monitor
Run 8   Event 0   DAQTime 0.00   Sun Feb 16 17:14:25 2003

In Event Block 100 events              In Time Region 1 seconds

Alpha \ n0    <    10      OK   > 4085   Alpha \ n0    <    10      OK   > 4085
Alpha<   10         0        0        0   Alpha<   10         0        0        0
Alpha OK            0        0        0   Alpha OK            0        0        0
Alpha> 8181         0        0        0   Alpha> 8181         0        0        0

Alpha \ n1    <    10      OK   > 4085   Alpha \ n1    <    10      OK   > 4085
Alpha<   10         0        0        0   Alpha<   10         0        0        0
Alpha OK            0        0        0   Alpha OK            0        0        0
Alpha> 8181         0        0        0   Alpha> 8181         0        0        0

Alpha \ n2    <    10      OK   > 4085   Alpha \ n2    <    10      OK   > 4085
Alpha<   10         0        0        0   Alpha<   10         0        0        0
Alpha OK            0        0        0   Alpha OK            0        0        0
Alpha> 8181         0        0        0   Alpha> 8181         0        0        0

Alpha \ n3    <    10      OK   > 4085   Alpha \ n3    <    10      OK   > 4085
Alpha<   10         0        0        0   Alpha<   10         0        0        0
Alpha OK            0        0        0   Alpha OK            0        0        0
Alpha> 8181         0        0        0   Alpha> 8181         0        0        0

Alpha \ all   <n_min      OK   >n_max   Alpha \ all   <n_min      OK   >n_max
Alpha<   10         0        0        0   Alpha<   10         0        0        0
Alpha OK            0        0        0   Alpha OK            0        0        0
Alpha> 8181         0        0        0   Alpha> 8181         0        0        0
```

**Figure 6 N-D Experiment – Parameter Tuning Monitor**

Program settings are displayed on the standard monitoring window (Figure 7) that contains sub-windows:
1. Main Setting – shows all parameters from configuration files
2. DAQ sequences – show all DAQ commands
3. History plot of event rate
4. Statistics Monitor - summarizes cross values from parameter tuning monitor
5. Event Monitor – Number of events and a single event monitor

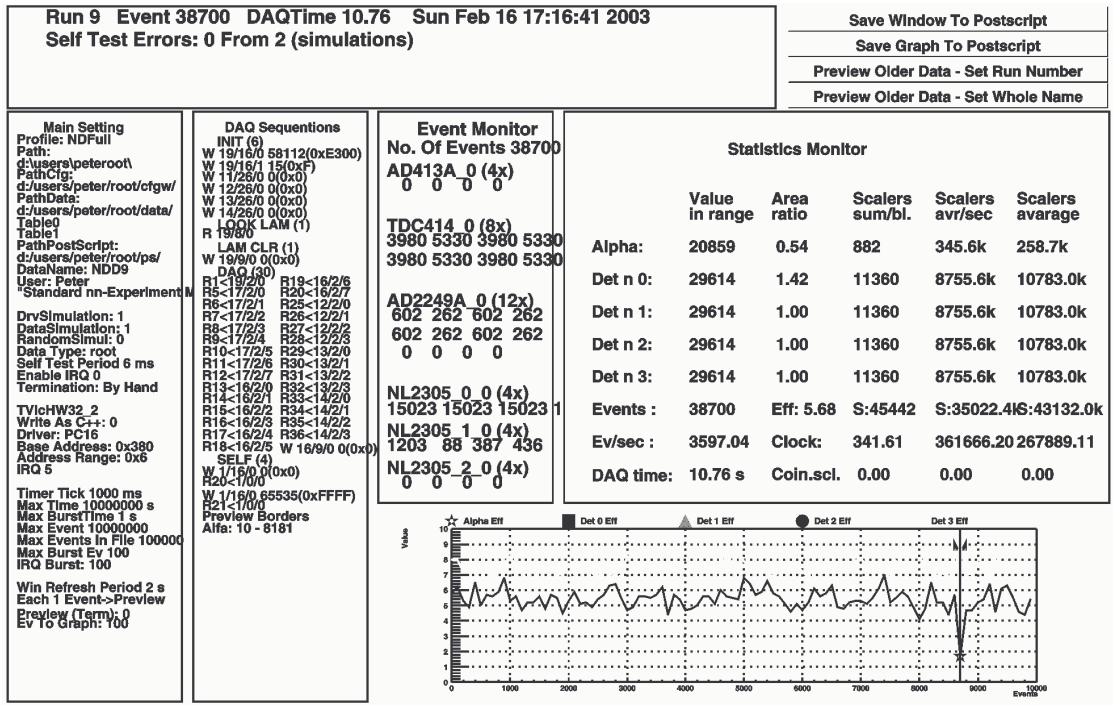

**Figure 7 N-D Experiment - Standard Monitoring Window**

Second set of monitors includes 1D and 2D histograms of several combinations of readout channels. Figure 8 shows 2D histograms, 1D histograms are shown in Figure 9. Figure 10 shows possibility to change windows arrangement and show most important plots in detail.

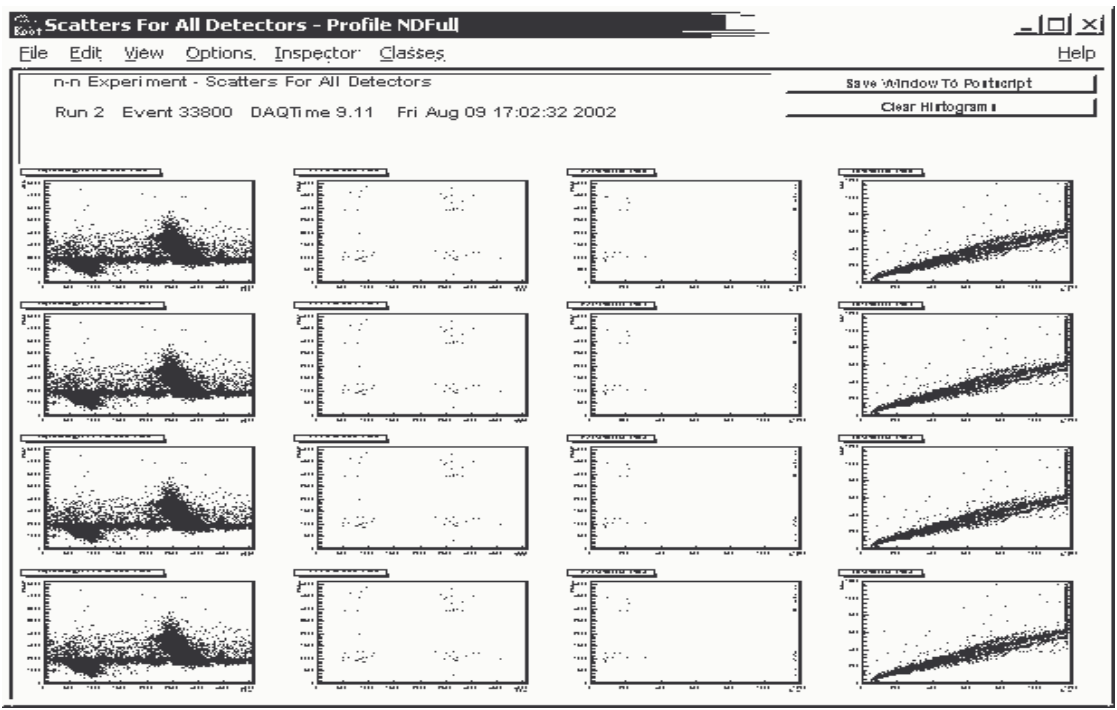

**Figure 8 N-D Experiment - Monitor of 2D Histograms**

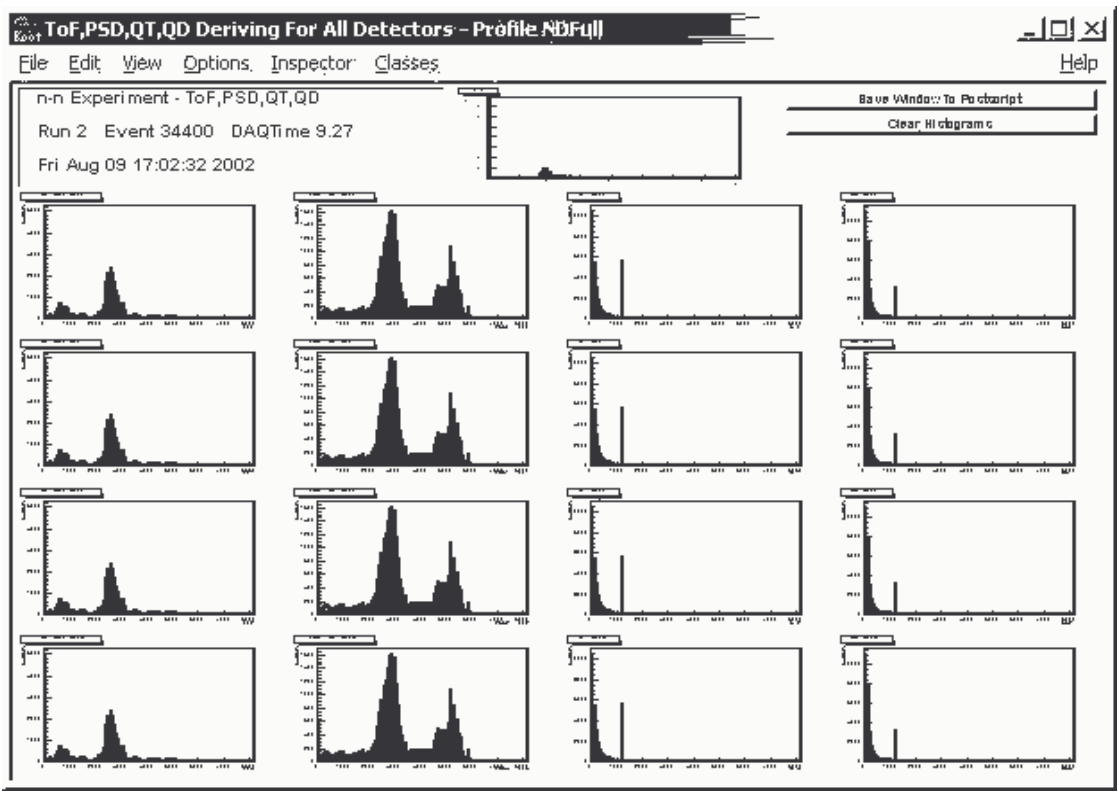

**Figure 9 N-D Experiment - Monitor of 1D Histograms**

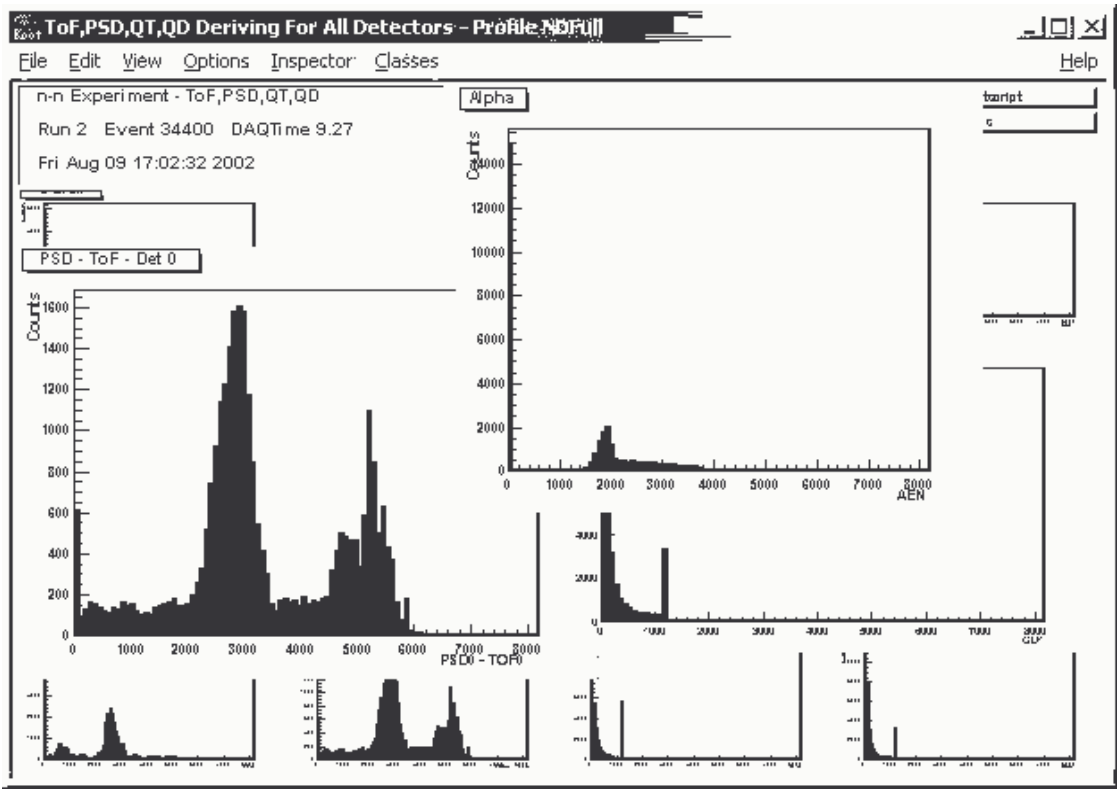

**Figure 10 N-D Experiment - Monitor of 1D Histograms - Details**

## b. Spectra Measurement

The spectra measurement includes a set of special setting and monitoring windows. The program supports also 2D spectra preview.
Some features of on-line spectra monitoring are:
1. Save source data event by event for later off-line analysis
2. On-line preview of 2D spectra

Some features of simple spectra analysis are:
1. Automatic comparison with tabular spectra in graphical regime
2. Automatic and manual peak detection in graphical regime, manual correction are possible
3. Automatic peak fitting and standard parameter calculations (net area, background subtraction)
4. Analysis in sub-range regions

Main monitoring windows with all DAQ spectra in 1D and 2D form are depicted in Figure 11. Individual spectra editing and analyzing is shown in Figure 12. Detail description of spectra measurement and analysis is in [6].

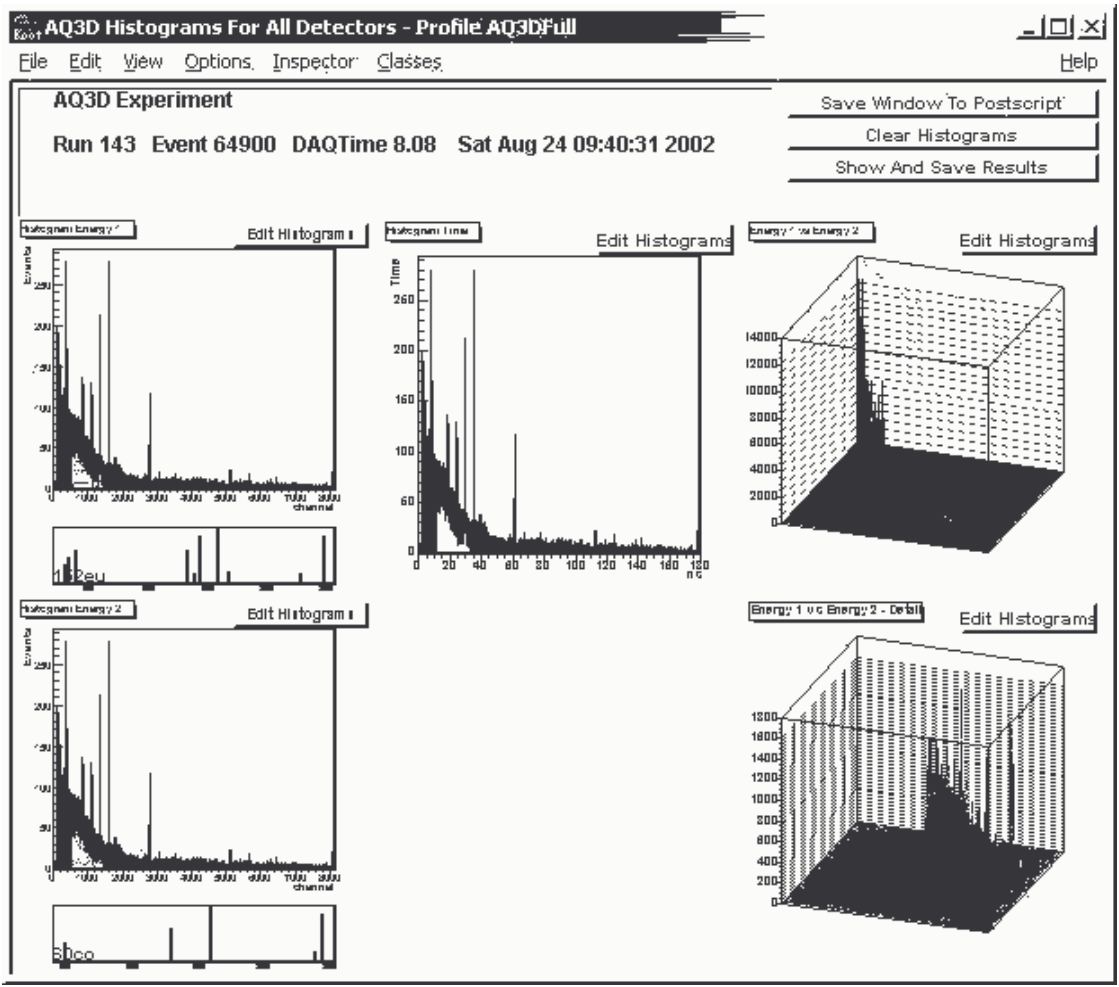

**Figure 11 Spectra Measurement - Main Monitoring Window With All DAQ Spectra**

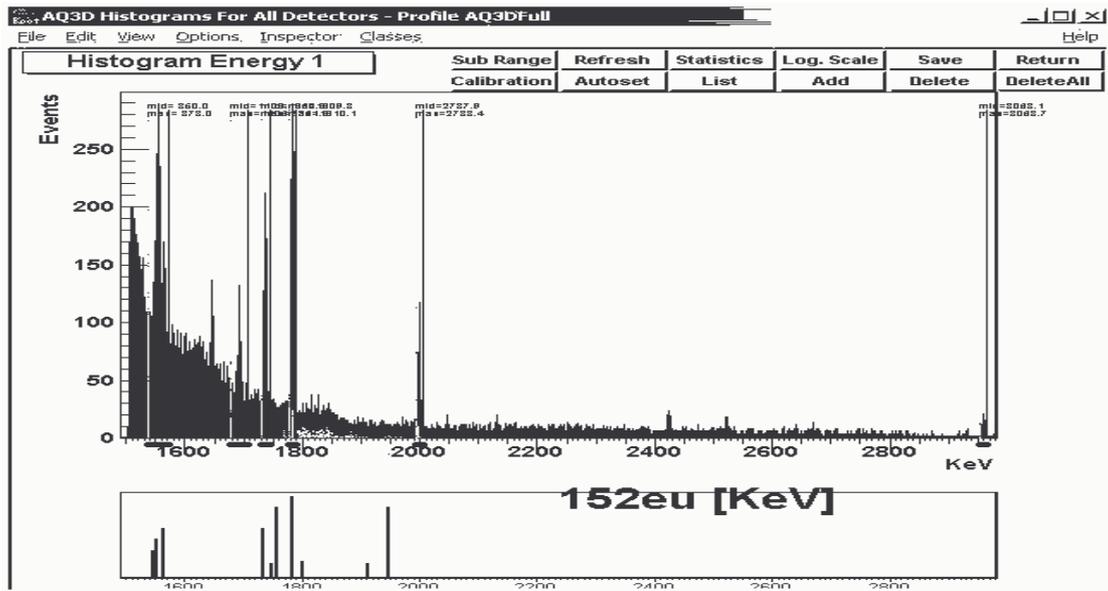

**Figure 12 Spectra Analysis – Spectrum Display**

## 6. Readout Speed Tests

Program was extensively tested on several types of computers, interface cards and with several versions of drivers. Its performance satisfies the requirements.

Two basic arrangements were tested: readout with CAMAC interface (Figure 13) and PC ISA digital I/O interface card PCD4848A (Figure 14) (readout of 48 TTL logical signals). A/D converter was used also as LAM signal producer. Three different CAMAC controllers were used.

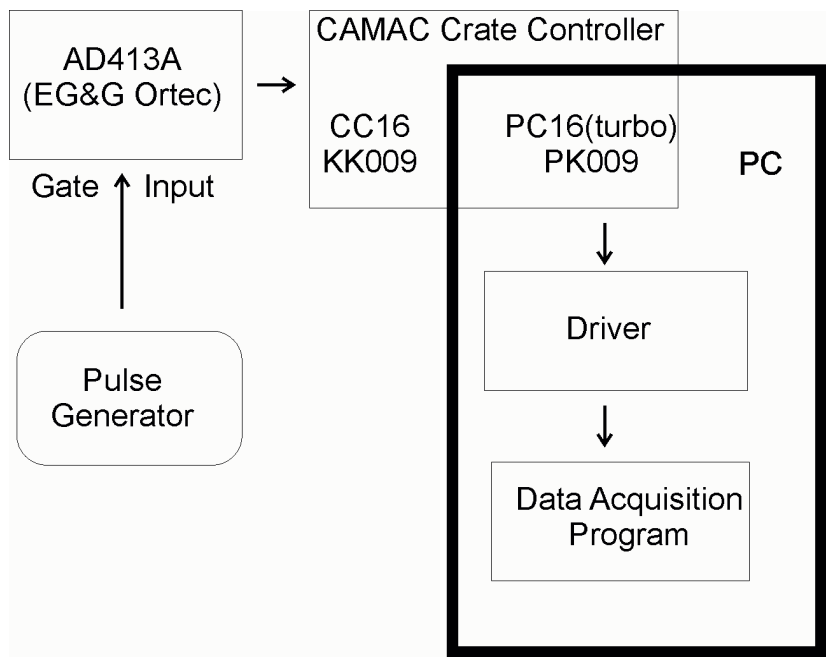

**Figure 13 CAMAC Devices Under Test**

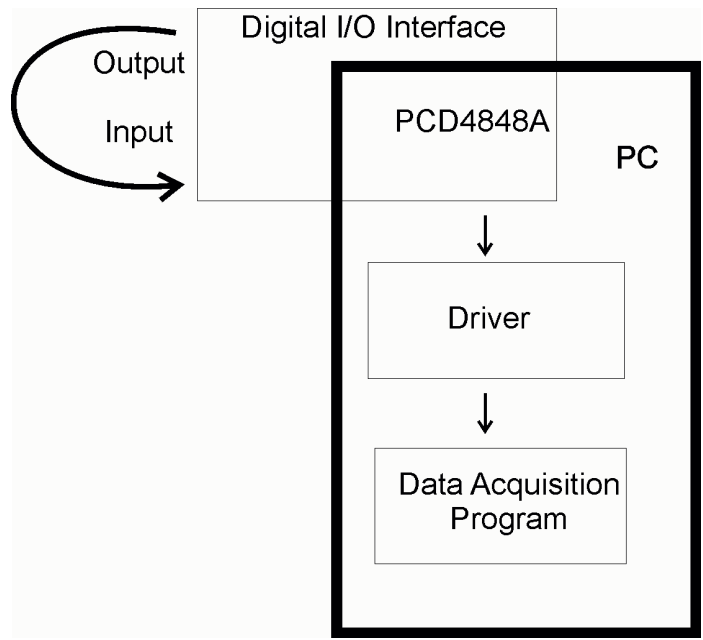

Figure 14 Digital I/O Device Under Test

**Tested PC interface cards:**
1. CAMAC crate controller: PC16 + CC16 (Wiener), Interrupt 5, port access, 8 CAMAC cycles per r/w
2. CAMAC crate controller: PC16turbo + CC16 (Wiener), Interrupt 5, High Speed regime (AT Bus regime decrease speed 1.3x), port access, 2 CAMAC cycles per r/w
3. CAMAC crate controller: PK009 + KK009 (Dubna), Interrupt 2, direct memory access (DMA), 1 CAMAC cycle per r/w
4. PC ISA interface: PCD4848A, Interrupt 5, port access – not CAMAC controller

Three types of PC ISA bus connection were tested. The simplest solution with direct addressing of driver memory space is possible only in DOS and partly in Windows 95/98. For Windows NT4.0/2000 there are several drivers available; we use TvicHW32 [7]. Used drivers have two different working modes: "soft" and "hard" access.

**Tested PC ISA bus drivers:**
TV2: TvicHW32 2.0 (1997, WinNT, W2K)
TV5: TvicHW32 5.0 (2001, WinNT, W2K)
"soft" access: feature of TvicHW driver - quick access to port without test of correct assignment to port
"hard" access: feature of TvicHW driver - slower access to port with test of correct assignment to port

**Testing sequences:**
read  (CAMAC: NF(0)A(0))
write  (CAMAC: NF(16)A(0))
cycle = write(0) + read + write(1) + read +count Number of errors (should be always zero)

Execution time was measured in microseconds for all tests: read, write and cycle. The results for different configurations used are listed in Table 1.

**Table 1 The results of DAQ speed for different testing configuration**

| Interface | Execution Time [μs] | | | Speed [kEvents / sec] | | PC Configuration |
|---|---|---|---|---|---|---|
| | Read | Write | Cycle | Read | Write | |
| PC16 | 23. | 23. | 92. | 40 | 40 | Pentium2 233MHz 128M RAM |
| PC16tur | 6.2 | 6.02 | 24.5 | 160 | 160 | Pentium2 233MHz 128M RAM |
| PK009 | 3.0 | 2.3 | 11.0 | 330 | 430 | Pentium2 233MHz 128M RAM |
| PCD4848A* | 2.2 | 2.2 | 9 | 450 | 450 | Pentium1 133MHz 128M RAM |
| PCD4848A* | 1.9 | 1.9 | 7.7 | 500 | 500 | AMD Duron(TM)Pro 600MHz, 256M RAM |

\* a more modern card produced in 2001, read/write BYTE, port access

From the test results we have collected the following conclusions:
- TvicHW32 driver: version 2.0 is approximately 10% faster than version 5.0.
- Get/set memory: "soft" access is approximately 10% faster than "hard" access.
- Get/set port: "soft" access is approximately 4 times faster than "hard" access – slightly depending on PC hardware.
- Speed slightly depended from PC power and from available memory
- Direct memory (DMA) operations are about 2 times faster than port operations (was not tested for DOS).
- Driver functions get/set memory are approximately 10% faster than standard C functions "memcpy"
- Standard C functions used in DOS (used standard C functions "movedata" (TurboC++ 1.1, 1990)) and in WinNT or W2K yield approximately the same read/write speed.
- WinNT and W2K yield approximately the same read/write speed.
- MSVC++ 6.0 application and ROOT + dynamic link library yield approximately the same read/write speed.
- Maximum speed of r/w is 330/430 kEvents / sec.
- Maximum speed for N-D experiment (about 40 r/w per event): 8/10 kEvents / sec (KK009, DOS / WinNT4.0 / Win2000) - results for PC2
- Modern PC ISA card PCD-4848A (made in 2001) works without problems in any PC with ISA bus. It seems that PK009, PC16 and PC16turbo (manufactured before 1998) are not very good for modern PC ISA bus - cards have too slow electronics.
- Larger memory (256 MB and more) increase multitasking and multithreading capabilities of DAQ: more monitoring windows and on-line histograms of multi parametric spectra, parallel off-line data processing, data transfers via network to data storage, parallel slow control routines.

**ISA Bus Properties Remarks**

Using crate controllers working on ISA bus is an archaic communication structure in modern PC. Several problems have been observed:
- It is impossible to work in emulated DOS under Windows because ISA bus is not connected directly to processor (Pentium2 and higher versions use PCI bus and north bridge for ISA bus connection, therefore different addressing [8])
- Possible problems with card installation: problem with BIOS conflicts, too much time needed for communication with board.
- Interrupt number – there is a possible conflict with mouse, network card, sound card, GPIB card, etc.
- Some computers (motherboard, processor, ISA architecture...) do not support communication with PC driver or support it only partially, perhaps due to an overly loose ISA Bus specification.

## 7. Discussion of Rate and Stability

The achieved DAQ speed can be increased only with difficulties. The main limit seems to come from the hardware used. The frequency of the CAMAC internal clock is 1 MHz and it involves a minimal reaction time higher than 1 microsecond. The DAQ speed was achieved 1.9 microseconds per ISA-bus-read/write; it is about 50% efficiency of the CAMAC system (1 microsecond bus cycle). The comparison with assembly language direct algorithms under DOS system for the same hardware configuration [9] shows a very similar efficiency of DAQ.

Possible rate fluctuations in current DAQ system are due to the use of Windows NT/2000 system, which runs system service programs at random intervals. The on-line visualization and service functions may also cause some rate fluctuations. A more stable rate can be achieved in DOS platform – of course without on-line monitoring support [9]. Rate fluctuations are occurring only at highest DAQ rates. Resulting fluctuations in dead time can be estimated using hardware scaler.

## 8. Conclusion

The presented DAQ program enables a high efficiency of data acquisition together with on-line monitoring calculations, histograms and scatter plots. Editing configuration files without recompilation of main DAQ program enables easy modifications of DAQ and monitoring conditions. Further supporting and servicing activities such as automatic creation of logbook, data backup/transfer, and messaging support about DAQ status are performed during the DAQ. The used ROOT platform supports flexibility of on-line monitoring functions and plots.
Detail information about program including downloadable version can be obtained at [6].

## 9. Acknowledgements

The project was granted by the *Grant Agency of the Czech Republic,* grant No. 202/00/0899. The authors want to express special thanks to Peter W. Phillips and Gareth F. Moorhead, whose SCTDAQ program was very inspiring for this project. Special thanks to Jan Jakůbek, Miroslav Morháč and Vladislav Matoušek for comments on WinNT drivers. We used a driver made by Victor Ishikeev to whom we are also very grateful. All critical comments from our colleagues, especially Vít Vorobel, Alex Tsvetkov and Michail Ivanov also helped us greatly.